\documentclass[aps,prl,twocolumn]{revtex4} 
\begin{document} 
{\bf \noindent Comment on ``Relevance of Cooperative Lattice 
Effects and Stress Fields in Phase-Separation Theories for CMR 
Manganites"} 
\\ 
\\ 
In a recent Letter, Burgy, Moreo, and Dagotto~\cite{Burgy04}
studied a modified random field Ising model (RFIM) to
show that structural disorder causes sub-micrometer scale
inhomogeneity of metallic and insulating domains in perovskite
manganites. The authors~\cite{Burgy04} attempted to include in their
spin model the ``cooperative effects" of  elasticity  to
resolve differences between their previous
models~\cite{Dagotto01} and experiments. In this Comment, we discuss
whether the spin models in Ref.~\cite{Burgy04}  are formulated adequately
for the problem at hand.
In particular,  we suggest that the
inhomogeneous features generated from the modified RFIM are the
consequence of the assumption that the metallic and insulating states
(modeled as spin up and down states) have  exactly the same energy
in the average field.
Moreover, we show that even if
one corrects for
this in the modified RFIM, a homogeneous phase is the result, 
which would not be sufficient
to explain the inhomogeneity in manganites.

In Ref.~\cite{Burgy04}, the modified RFIM Hamiltonian $H=-J 
\sum_{<i,j>} s_i s_j -\Delta \sum_{i,j} h_i s_j / d^{\alpha}_{ij}$ 
has been considered, where $s_i$ are Ising variables, $J$ is the 
ferromagnetic coupling, $\Delta$ is the disorder strength, 
$h_i$ are random perturbations,
and 
$d_{ij}$ is the distance between lattice sites $i$ and $j$. By 
defining $\tilde{h}_j=\sum_{i}h_i/d^{\alpha}_{ij}$, the 
Hamiltonian is expressed as $H=-J \sum_{<i,j>} s_i s_j 
-\Delta \sum_j \tilde{h}_j s_j$. It represents the 
situation in which random perturbations $h_i$ have a non-local 
interaction with spins $s_j$ with strength that decays as
$1/d^{\alpha}_{ij}$. The authors of Ref.~\cite{Burgy04} introduced 
this non-local interaction to mimic the long-range nature of 
elastic fields~\cite{Shenoy99}. 

We first note that the exponent values used in
showing the domains from the simulations  are $\alpha = \infty 
$, 3, and 1  on a 2-dimensional (2D)
lattice (Fig.~2 Bottom panel of Ref.~\cite{Burgy04}). However,  
the long-range elastic interaction in 2D decays as $\alpha $ = 2
(Ref.~\cite{Shenoy99}).  More importantly, the average of the random perturbation 
has been chosen to be zero (i.e., $<h>_{space}$ = 0), which would 
imply that the metallic and insulating states, modeled as spin up 
and down, are exactly degenerate in the average perturbation.  In
Ref.~\cite{Burgy04}, it is suggested that the random distribution 
of rare earth and alkali metal ions, which have different ionic 
sizes, gives rise to the structural random perturbations represented as 
$h_i$. Our point is that the average of $h_i$ should correspond to the 
average ionic size at rare earth and alkali metal sites, and  the 
spin up and down states in the modified RFIM should, in general, have different 
energies in the average perturbation due to the lack of symmetry between 
metallic and insulating states. In other words, $<h>_{space}$ should have 
a non-zero value, similar to the non-zero $h$ field in Ising 
models applied to binary alloys or lattice gases~\cite{Ma}. 
 
If  we consider $\alpha$ = 2 (3 in 3D)
and $<h>_{space} \neq 0$, then because of the long range nature of $1/d^2$ 
in 2D ($1/d^3$ in 3D), the net field $\tilde{h}$ at any site 
diverges logarithmically with the sign of $<h>_{space}$, which means that 
the system should be homogeneous -- contrary to the desired goal of the
authors of Ref.~\cite{Burgy04}. The inconsistency demonstrates 
that the results of the spin model in Ref.~\cite{Burgy04}
have to be considered carefully in interpreting the role of
elasticity and structural randomness for the inhomogeneity seen in 
perovskite manganites. 
 
\vspace{0.5cm} 
 
\noindent K. H. Ahn\cite{Ahn} and T. Lookman 
 
 
{\small Theoretical Division 
 
 
Los Alamos National Laboratory 
 
 
Los Alamos, New Mexico 87545} 
 
\vspace{0.5cm} 
 
 
 
{\noindent \small PACS numbers: 75.47.Lx, 75.10.-b, 75.30.Kz }

\end{document}